\def\compileforpublish{1}
\def\isaccepted{1}

\documentclass[conference]{ieeeconf}  %

\IEEEoverridecommandlockouts                              %

\usepackage{tabularx,calc}
\usepackage{color}
\usepackage{tubscolors}

\usepackage{tikz}
\usetikzlibrary{%
	shapes,
	automata,
	arrows,
	matrix,
	backgrounds,
	fit,
	patterns,
	decorations.markings, 
	svg.path, 
	shapes.multipart, 
	external, 
	shadows, 
	positioning, 
	calc, 
	decorations, 
	intersections, 
	angles, 
	quotes,
	decorations.pathmorphing, 
	arrows.meta, 
	patterns, 
	decorations.pathreplacing, 
	external,
	snakes,
	spy%
}

\usepackage{tkz-euclide}

\usetkzobj{all}
\usepackage{pgfplots}
\usepackage{pgfplotstable}
\usepgfplotslibrary{groupplots}
\usepgfplotslibrary{statistics}
\usepackage{tikz-3dplot}

\usepackage{siunitx}
\usepackage[%
			bibstyle=ieee,
			citestyle=numeric-comp,
			backend=bibtex,
			minnames=1,
			maxcitenames=2,
			maxbibnames=99,
			doi=true,
			isbn=false,
			url=false,
			natbib=true]
			{biblatex} 
\usepackage[caption=false, font=footnotesize]{subfig}
\usepackage{hyperref}

\usepackage{ifthen}
\usepackage{lipsum}

\usepackage{xstring}
\usepackage[abspath]{currfile}

\usepackage{bm} 		%

\usepackage{amssymb} 	%
\usepackage{amsmath}
\usepackage{siunitx}
\usepackage{graphicx}
\usepackage{bm} 		%
\usepackage{booktabs} 	%
\usepackage{multirow}   %
\usepackage{booktabs} 	%
\usepackage{colortbl}
\usepackage{array}

\usepackage{microtype}

\AtBeginDocument{} 
\AtBeginDocument{} 

\newcolumntype{P}[1]{>{\centering\arraybackslash}p{#1}}
\makeatletter
\newcounter{IEEE@bibentries}
\renewcommand\IEEEtriggeratref[1]{%
	\renewbibmacro{finentry}{%
		\stepcounter{IEEE@bibentries}%
		\ifthenelse{\equal{\value{IEEE@bibentries}}{#1}}
		{\finentry\@IEEEtriggercmd}
		{\finentry}%
	}%
}
\makeatother

\ifx \isaccepted \undefined
\newcommand\copyrighttext{%
	\footnotesize \centering This work has been submitted to the IEEE for possible publication.\\ Copyright may be transferred without notice, after which this version may no longer be accessible.}
\else
\newcommand\copyrighttext{%
	\footnotesize \parbox[t]{.11\textwidth}{\copyright{} \the\year~IEEE.} \parbox[t]{.88\textwidth}{Personal use of this material is permitted. Permission from IEEE must be obtained for all other uses, in any current or future media, including reprinting/republishing this material for advertising or promotional purposes, creating new collective works, for resale or redistribution to servers or lists, or reuse of any copyrighted component of this work in other works.}}
\fi
\newcommand\copyrightnotice{%
	\ifx \compileforpublish \undefined
	\else
	\begin{tikzpicture}[remember picture,overlay]
	\node[anchor=south,yshift=10.5pt] at (current page.south) {\parbox{\dimexpr\textwidth-\fboxsep-\fboxrule\relax}{\copyrighttext}};
	\end{tikzpicture}%
	\fi
}

\newcommand{\todo}[2][]{%
	\ifthenelse{\equal{#1}{}}{\def\mynamestring{}\def\mycolor{red}}{\def\mynamestring{\lbrack\uppercase{#1}\rbrack}\def\mycolor{#1}}
	\emph{\textcolor{\mycolor}{(TODO\mynamestring: #2)}}%
}

\tikzset{external/system call={pdflatex -enable-write18 -halt-on-error
		-interaction=batchmode -jobname "\image" "\texsource" -aux-directory=\currfileabsdir/auxfiles_tikz}}

\graphicspath{{figures/images/tikz/}{figures/images/raster/}{figures/images/vector/}}

\newcommand{\getimgpath}{%
	\StrBehind{\currfiledir}{/}[\imgbase]%
}

\newcommand{\myincludegraphics}[2][]{%
	\getimgpath%
	\IfSubStr*{\currfileext}{tikz}{\StrBehind*{\imgbase}{images/tikz/}[\imgbase]}{}
	\includegraphics[#1]{\imgbase#2}
}

\verbtocs{\bs} |\|

\newcommand{\getfilepathbs}{%
	\StrSubstitute{\currfiledir}{\bs}{/}[\thefilepathbs]%
}

\newcommand{\getrelimgpath}{%
	\getfilepathbs
	\StrBehind*{\thefilepathbs}{content/}[\relimgpath]%
}

\newcommand{\inputtikz}[2][]{%
	\getrelimgpath%
	\ifx&#1&%
	\tikzpicturedependsonfile{#1}%
	\fi%
	\StrBefore{\detokenize{#2}}{.}[\filename]%
	\ifx&\externalize&%
		\tikzsetnextfilename{\filename}%
	\fi%
	\input{figures/images/tikz/\relimgpath\detokenize{#2}}%
	\gdef\externalize{}
}
\newcolumntype{L}[1]{>{\raggedright\arraybackslash}p{#1}} %

\newcommand{\MATH}{\textsc{M\footnotesize ATHEMATICA\normalfont}}

\renewcommand{\vec}[1]{\underline{#1}}
\def\mat#1{\underline{\underline{#1}}}

\newcommand{\phidot}{\dot{\varphi}}

\newcommand{\psidot}{\dot{\psi}}
\newcommand{\psiddot}{\ddot{\psi}}
\newcommand{\thetadot}{\dot{\theta}}

\definecolor{cr}{RGB}{102,180,211}
\definecolor{mn}{RGB}{255,127,0}
\definecolor{rs}{RGB}{190,0,80}

\input{shapes.tex}
\tikzstyle{shadedBlue} = [top color=tuDarkBlue20, bottom color=tuDarkBlue40, draw=tuDarkBlue80, thick]%
\tikzstyle{shadedRed} = [top color=tuRed20, bottom color=tuRed40, draw=tuRed80, thick]%
\tikzstyle{shadedOrange}  = [top color=white, bottom color=tuOrange40,  draw=tuOrange100, thick]%
\tikzstyle{shadedGreen}  = [top color=white, bottom color=tuGreen40,  draw=tuGreen100, thick]%
\tikzstyle{shadedYellow} = [top color=white, bottom color=tuYellow40,  draw=tuYellow100, thick]%
\tikzstyle{shadedGray} = [top color=white, bottom color=tuGray20, draw=tuGray60, thick]%
\tikzstyle{shadedGrayLight} = [top color=tuBlack!5, bottom color=tuBlack!10, draw=tuBlack!30, thick]%
\tikzstyle{shadow} = [drop shadow={opacity=.5,shadow xshift=.3ex,shadow yshift=-.3ex}]%
\tikzstyle{triangle} = [isosceles triangle,isosceles triangle stretches]%
\tikzstyle{label-it} = [font=\itshape]%
\tikzstyle{block} = [draw, shadedBlue, rectangle, rounded corners, minimum height=2em, minimum width=5em]%
\tikzstyle{smallblock} = [draw, shadedBlue, rectangle, rounded corners, minimum height=1em, minimum width=2em,shadow]%
\tikzstyle{outerblock} = [draw, shadedGrayLight, draw=tuGray80, rectangle, rounded corners, minimum height=2em, minimum width=5em,shadow]%
\tikzstyle{bubble} = [fill=black,shadow,circle,draw=black,inner sep=0pt,minimum size=5pt]%
\tikzstyle{memory} = [cylinder, shape border rotate=90, aspect=.4, shadedGrayLight, minimum width=5em, shadow]%
\tikzstyle{inheritArrow} = [-open triangle 60,thick]%
\tikzstyle{kompArrow}    = [diamond-,thick]%
\tikzstyle{flowDecision} = [diamond, draw, shadedRed, text badly centered, inner sep=0pt,shadow]%
\tikzstyle{flowBlock} = [rectangle, draw, shadedBlue, text centered, rounded corners, minimum height=2em,shadow]%
\tikzstyle{bgBox} = [rectangle, draw, shadedGrayLight, text centered, rounded corners=3mm, shadow, inner sep=10pt]%
\tikzstyle{blockarrow} = [draw, thick, single arrow, minimum height=3em]%

\tikzstyle {archblock} = [outerblock, minimum height=4em, align=center, minimum width=12em, font=\sf]
\tikzstyle {slim} = [minimum width=6em]
\tikzstyle {verticalblock} = [archblock, rotate = 90, archblock, minimum width=16.5em, minimum height=2em]
\tikzstyle {sensors} = [rectangle split parts=2, rectangle split horizontal]

\tikzstyle {sup} = [yshift=0.5cm]
\tikzstyle {sdown} = [yshift=-0.5cm]
\tikzstyle {sleft} = [xshift=-0.5cm]
\tikzstyle {sright} = [xshift=0.5cm]

\tikzstyle {arrow} = [very thick, -stealth']

\tikzstyle {seperator} = [thick, dashed, lightgray]
\tikzstyle {arrow} = [-stealth', thick]
\tikzstyle {rarrow} = [arrow, stealth'-]

\makeatletter
\tikzset{reset preactions/.code={\def\tikz@preactions{}}}
\makeatother

\tikzstyle {disable} = [reset preactions, draw=none, fill=none, top color=white, bottom color=white]
\tikzstyle {header} = [disable, font=\sf]

\tikzset{mysplit/.style={rectangle split, rectangle split parts=2, rectangle split draw splits=false, inner sep=2.2ex,
		rectangle split horizontal,minimum width=5.5em},
	textstyle/.style={text height=1.5ex,text depth=.25ex}}

\title{\LARGE \bf
	Sensitivity Analysis for Vehicle Dynamics Models -- An Approach to Model Quality Assessment for Automated Vehicles
}

\author{Marcus Nolte$^{1}$, Richard Schubert$^{1}$, Cordula Reisch$^{2}$, and Markus Maurer$^{1}$%
	\thanks{\hspace{-1em}$^{1}$Institute of Control Engineering, TU Braunschweig, Germany
		{\tt\small \{nolte, maurer\}@ifr.ing.tu-bs.de}\vspace{0.3em}\newline%
		$^{2}$Institute for Partial Differential Equations, TU Braunschweig, Germany
		{\tt\small \{c.reisch\}@tu-braunschweig.de}}%
}

\begin{document}
	\renewcommand{\sectionautorefname}{Section}%
	\renewcommand{\figureautorefname}{Fig.}%
	\maketitle%
	\thispagestyle{empty}%
	\pagestyle{empty}%
	\begin{abstract}%
		Model-based approaches have become increasingly popular in the domain of automated driving.
This includes runtime algorithms, such as Model Predictive Control, as well as formal and simulative approaches for the verification of automated vehicle functions.
With this trend, the quality of models becomes crucial for automated vehicle safety.
Established tools from model theory which can be applied to assure model quality are uncertainty and sensitivity analysis \cite{reuter2011}.

In this paper, we conduct sensitivity analyses for a single and double track vehicle dynamics model to gain insights about the models' behavior under different operating conditions.
We compare the models, point out the most important findings regarding the obtained parameters sensitivities, and provide examples of possible applications of the gained insights.
	\end{abstract}%
	\copyrightnotice
	\section{Introduction}
\label{sec:intro}

Recent developments in the field of automated driving show strong efforts to transition from simple demonstrations to a large-scale industrialization of SAE Level 4+ vehicles.
Many companies have announced mobility services based on Level 4+ vehicles \cite{SAE2014} within the next years.
Waymo has e.g. just started to offer SAE Level 4 shuttle services in Phoenix, AZ \cite{hawkins2019}.
While Waymo's Operational Design Domain (ODD) is not overly complex so far, e.g. in terms of traffic density, other companies such as Zoox are planning on releasing Level 4 shuttles in Downtown San Francisco in 2020 \cite{coates2019}.

A key challenge for Level 4+ systems is having to cope with malfunctions without any human intervention.
Hence, monitoring of the overall system health, the quality of the executed function and the ODD boundaries becomes crucial for safe operation.
In this respect, the system must not only be able to detect possible faults, but it must also be able to enter a risk minimal state, even in a degraded condition.
While monitoring and the representation of system models have been a key concept for autonomous systems design for several decades \cite{antsaklis1989}, both are complex tasks, particularly for perception systems which are driven by machine-learning-based algorithms.
At the same time, both remain challenging tasks in the planning and control domain, as well.

With increased computational power, recent years have shown a growing trend toward model-based (e.g. Model Predictive Control) optimization approaches for trajectory planning and vehicle control.
While these approaches show promising results, they heavily depend on the quality of the underlying models. 
In the best case, model mismatch causes sub-optimal system behavior \cite[1]{thangavel2018}.
In the worst case, model mismatch causes vehicle behavior deviating form originally formulated safety constraints \cite[1]{thangavel2018}, as their derivation is often based on the applied dynamics models.

With regard to monitoring, it is thus desirable to gain information about model validity.
This is true not only during the development of algorithms, but also crucial at runtime when the targeted system requires a certain degree of autonomy \cite[332]{antsaklis1989}.
At the same time, when designing model-based algorithms, a careful consideration of the models' strengths and weaknesses is highly safety relevant.

A typical example of such strength and weaknesses are the assumptions made for the linear single-track model, which is often used in literature.
A core assumption for the derivation is constant velocity to eliminate non-linear state dependencies.
However, single-track models are often used in a parameter varying fashion for describing lateral vehicle dynamics.
With these simplifications, performance indicators with respect to model quality are needed.
This could e.g. cause the question of how far velocity and acceleration may change, before causing insufficient model accuracy.
- I.e. the sensitivity of the model with respect to the velocity parameter.

In literature, there is a number of studies dealing with experimental model-validation \cite{polack2018, matute2019}.
Some studies have explicitly addressed the question of model validity for single track models \cite{polack2017,altche2017}.
Finally, there are some approaches to sensitivity analysis for single-track \cite{jang1997, hamza:tel-01347105} and the Pacejka tire model \cite{hamza:tel-01347105}.
However, fundamental approaches based on model-theoretic analysis, such as sensitivity analysis have to the knowledge of the authors currently not been applied driven by the requirements for automated driving.

Hence, this paper contains the following contributions:
We provide an overview of approaches for sensitivity analysis and the benefits of model-theoretic approaches in the context of safety for automated vehicles.
We review a sensitivity analysis for a linear dynamic single-track model and extend it for rear steering. 
We evaluate it in different driving situations and compare to the findings of \cite{polack2017} and \cite{hamza:tel-01347105}.
Finally, we conduct a sensitivity analysis for a non-linear double track model and point out the main findings with respect to the sensitivities.

The paper is structured as follows:
\autoref{sec:literature} presents related work regarding sensitivity analyses and the assessment of vehicle dynamics models. 
\autoref{sec:sensitivity} gives a model-theoretic motivation for conducting model-theoretic analyses.
The section presents fundamental of sensitivity analysis and motivates a model-theoretic discussion from the perspective of automated driving.
\autoref{sec:application} applies the theoretic concepts to a single- and a double-track model, before \autoref{sec:results} presents the results and \autoref{sec:conclusion} concludes the paper.
	\section{Related Work}
\label{sec:literature}

Methods for sensitivity analysis can be separated into \emph{local} and \emph{global} methods \cite{ye2017}.
Local methods can be applied if a nominal estimate for model parameters is available.
By partial derivation of states and outputs with respect to a parameter (cf. \autoref{sec:sensitivity}), its influence on a given state or output can be determined.
Global methods in contrast, can be used if a nominal estimate of the paramters is not available.
These methods work over entire parameter ranges.
Global approaches are either based on multiple local sensitivity analyses or variance-based methods, requiring Monte Carlo sampling to explore the parameter space, what makes them computationally challenging.
For the vehicle dynamics models at hand, we would like to evaluate local sensitivities, as initial estimates of the physical parameters are available.

As mathematical models are always simplifications of real processes, one of the most important model-theoretic questions is how to measure the quality of a model. 
From a model-theoretic perspective, uncertainty and sensitivity analysis are established mathematical tools assisting the modeling process.

In this context, \citeauthor{reuter2011} describe a general framework to ensure model accuracy \cite{reuter2011}. 
After determining parameter values by data-driven approaches, they propose to perform a sensitivity analysis for judging the dependency of the model on a certain parameter value. 
In a next step, they point out the importance of model validation using independent data, which has not been used for identification.
Finally, they demand proper documentation of the results obtained during sensitivity analysis and validation. 

As described in \autoref{sec:intro}, the aforementioned methods are still rarely applied in the context of automated driving, despite high (safety) requirements which apply to the systems.
\citeauthor{polack2017} present studies on the fitness of a single track model for motion-planning applications:
In \cite{polack2017}, they explicitly motivate the need for consistent models between planning and control modules in layered architectures.
They particularly state that this is the case when models are applied at different levels of abstraction, e.g. due to computational power demands.
In this context, they compare a kinematic bicycle model to a 9 DOF model at different (constant) curvatures and velocities.
Finally, they derive a maximum lateral acceleration of $0.5 g$ for sufficient consistency between both models from their simulations.

In \cite{polack2018}, the same authors apply these findings for the implementation of an MPC-based local trajectory planner and an underlying trajectory controller.
The framework is designed to avoid the derived critical lateral acceleration limit by introducing it as a constraint in the optimization problem.
\citeauthor{matute2019} implement a similar MPC framework, also based on the findings of \cite{altche2017}, while experimentally validating their model parameters.

\citeauthor{hamza:tel-01347105} conducts a sensitivity analysis for a linear single-track vehicle model \cite{hamza:tel-01347105}.
The author states the importance of accurate models for vehicle stabilization algorithms and the need to cope with parameter uncertainty.
The sensitivity analysis is complemented by a global sensitivity analysis for a Pacejka tire model, as well as a method for parameter fine tuning of the tire model.
However, the analysis of the single track model is only qualitatively evaluated in a simple double lane change maneuver.
The studies provide no perspective toward automated driving.
The same holds true for \cite{jang1997}.
While the authors present a sensitivity analysis for a single track model, the described results are inconsistent, as sensitivities for the yaw rate with respect to the vehicle mass are found, while the yaw rate only depends on the moment of inertia.

The importance of model accuracy is also pointed out from a control-theoretic perspective by \cite{thangavel2018, simkoff2017}, who present MPC approaches with integrated handling of model-plant mismatch.

In summary, questions of model quality and the handling of model uncertainty are still subject to current research, despite the availability of long established mathematical methods.
Particularly for safety-critical applications it thus becomes crucial to apply those tools to gain detailed insight about the applied models.
	\section{Sensitivity Analysis}
\label{sec:sensitivity}

From a very abstract model-theoretic point of view, as taken by \cite{langemann2018}, we can assume the world to be deterministic in its underlying mechanisms.
The mechanisms are generally unknown, even if we find situations, where a physical model approximates the real world sufficiently well.
By capturing the unknowns in approximate models, the deterministic nature of the models can also turn probabilistic. 
Let the system describing the world be $\dot{Y}(t)= G(Y(t))$ with $Y(0)=Y_0$, where $Y$ is the vector of the system state and $G$ contains all mechanisms. 
In this framework, a model is describing a part of the constructed reality. 
Let the model be described by $\dot{X}(t)= F_c(X(t))$ with $X(0)=X_0$, depending on parameters $c$. 
Due to simplifications and generalizations, the state variables $Y$ and $X$ are usually not the same and typically not of the same size. 

Hence, it is normally not possible to measure the variables $X$ or even $Y$ directly. 
We name the measured quantities $W=W(t)$.
The modeling process is now driven by the hope that there is a connection $\psi$ with $W(t) \approx \psi ( Y(t))$.
In this general setting, we use the fluxes $\Gamma=\Gamma(t)$ and $\Phi_c=\Phi_c (t)$ as a description of the differential equations given by $G$ or $F$, respectively. 
In the latter, the system $\dot{X}(t)=F_c(X(t))$ is depending on parameters~$c$.

\begin{figure}%
	\centering%
	\begin{minipage}{.25\textwidth}%
		\myincludegraphics{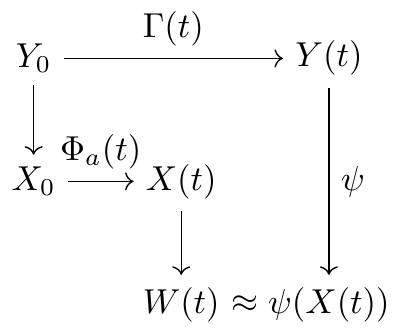}%
	\end{minipage}%
	\caption{%
		Theoretical modeling framework with a real world model with state $Y=Y(t)$, a model state $X=X(t)$ and measured data $W(t)$.%
	}%
\end{figure}  

This theoretical modeling framework highlights the fact, that it is oftentimes impossible to measure the parameters directly, which introduces additional uncertainty into the modeling process. 
At the same time, the model $\dot{X}(t)=F_c(X(t)) = F_c(x_1(t), x_2(t), \dots, x_n(t),  c_1, c_2,\dots, c_m)$ and therefore the model results $X_c(t)$ are heavily depending on correctness of the model's parameters. 

A sensitivity analysis provides a possibility for studying the dependency of the model results on the parameters. 

In the following, we follow the notations of \cite{dickinson1976} and \cite {perko2001}.
We write the sensitivities of the i-th state with respect to changes of the k-th model parameter as
\begin{equation}%
Z_{i,k} = \frac{\partial x_i}{\partial c_k},\qquad i = 1,\dots, n\quad k = 1,\dots, m.%
\label{eq:sens}%
\end{equation}%
The time derivative of (\ref{eq:sens}) allows to directly use the time derivatives of the system's states for a calculation of the sensitivities alongside the model's states.
Hence the presented method is also often referred to as the \emph{Direct Method} for sensitivity analysis.
Derivation with respect to time, applying the chain rule and changing the order of differentiation yields
\begin{equation}%
\dot{Z}_{i,k} = \frac{\partial f_i}{\partial c_k} + 
\left ( \sum_{j=1}^N \frac{\partial f_i}{\partial x_j}\cdot Z_{j,k} \right ).
\end{equation}%
Rewriting the sum as the product of the original system's Jacobian $\mat{J} \in \mathbb{R}^{n\times m}$ and the vector of sensitivities $\vec Z \in \mathbb{R}^m$ yields the \emph{sensitivity system}
\begin{equation}%
\vec{\dot{Z}}=\vec{f_c}+\mat{J}\,\vec{Z}
\end{equation}%
which describes linear system dynamics for the system's sensitivities with a parameter varying transition matrix.
$f_c$ is an input vector to the sensitivity system holding the partial derivative of each state function with respect to each parameter
\begin{equation}%
\vec{f_c} := \left ( \frac{\partial f_1}{\partial \vec c}, \: ...\:, \frac{\partial f_n}{\partial \vec c} \right )^T.
\end{equation}%

Since the sensitivity system is an ordinary differential equation (ODE) system in itself, a note regarding its initial values for the vector of sensitivities is required:
As described in \cite{dickinson1976}, the initial value is defined as
\begin{equation}%
Z_{i,k}(0) = \lim_{\Delta c_k\to 0}\left \{ \frac{x_i(c_k+\Delta c_k, 0) - x_i(c_k, 0)}{\Delta c_k}\right \}%
\end{equation}%
with an arbitrarily small change in the $k$-th parameter $\Delta c_k$.
As no parameter is an initial value of the differential equation system, the initial value, according to \cite{dickinson1976}, can be set to
\begin{equation}%
Z_{i,k}(0) = 0,\:\forall i,k.%
\end{equation}%

\subsection{Relevance for automated driving}
\label{sec:relevance}

The model-theoretic motivation above can be easily transferred to the domain of automated driving.
The systems have to navigate a highly complex environment (real world) such that high-quality models become a key asset to a variety of applications in the field.
Whether it is model-based planning and control as described in \autoref{sec:intro}, or validation and verification approaches:
Safety guarantees, which must hold in a real-world application, can only be trusted, if the (formal) verification process is based on sufficiently accurate models.

Verification and validation is a question of the development process:
For safety verification of planning and control algorithms, it is e.g. desirable to make statements about worst-case input-output relations of (controlled) vehicle dynamics models, e.g. by applying reachability analysis \cite{klischat2019,rizaldi2018}.
When applied with proper system knowledge, such approaches can yield valuable formal proof whether safety constraints can be adhered to.
While this is a powerful tool, it should be obvious, that the value of such proofs only holds with respect to modeling assumptions and simplifications made during the verification process.
The same is true for control-quality guarantees which are given at design time or the results obtained by simulative validation and verification.

Due to the inevitable presence of uncertainty in all of the application domains mentioned above, robust and stochastic methods have been developed, such as stochastic reachability analysis or robust / stochastic MPC.
However, while these approaches are again powerful in their theoretic capabilities, their performance is extremely sensitive to a precise quantification of uncertainty.
In practice, the theoretical advantages of such approaches are nullified, if e.g. parameter distributions or robustness parameters are empirically tuned rather than carefully analyzed and applied.
This is an area of application where uncertainty- and sensitivity analysis provide a profound theoretical framework for an adequate formulation of stochastic or robust planning or control algorithms.

Apart from an application in probabilistic settings, sensitivity analysis can also provide valuable guidance for parameter identification and system design in general.
Sensitivities describe the impact of parameters on system states.
From a development perspective, they can hence be used to determine sensitive parameters, which must be identified with high accuracy.
In addition, sensitivity information can also be used to perform model reduction, e.g. by eliminating insensitive parameters from the system's state equations.
Under real-time aspects, this yields a profound argument for making trade offs between model complexity and execution times of model-based algorithms.
While making these trade offs, a deeper model-theoretic analysis allows to establish relations between the derived model (an example will be given in Section \ref{sec:results}).
These relations provide additional semantic information e.g. about how different vehicle dynamics models interact, also regarding consistency, as demanded by \citeauthor{polack2017} \cite{polack2017, polack2018}.

Regarding, parameter sensitivities can also provide valuable input for fault isolation.
Analyzing residuals between predicted and actual vehicle behavior can e.g. yield hints that the quality of model-based algorithms is degrading.
By definition (\ref{eq:sens}), an additional sensitivity analysis at runtime can yield complimentary information, which parameters are responsible for possible deviations in a given situation (cf. \autoref{sec:results}).
This in turn contributes to the autonomy of the system, as discussed in \cite{antsaklis1989}.
The additional information enhances the demanded representation of models beyond a pure state-space representation and can provide hints towards model validity.
	\section{Application to vehicle dynamics models}
\label{sec:application}

\begin{figure}
	\centering
		\includegraphics[width=.95\columnwidth]{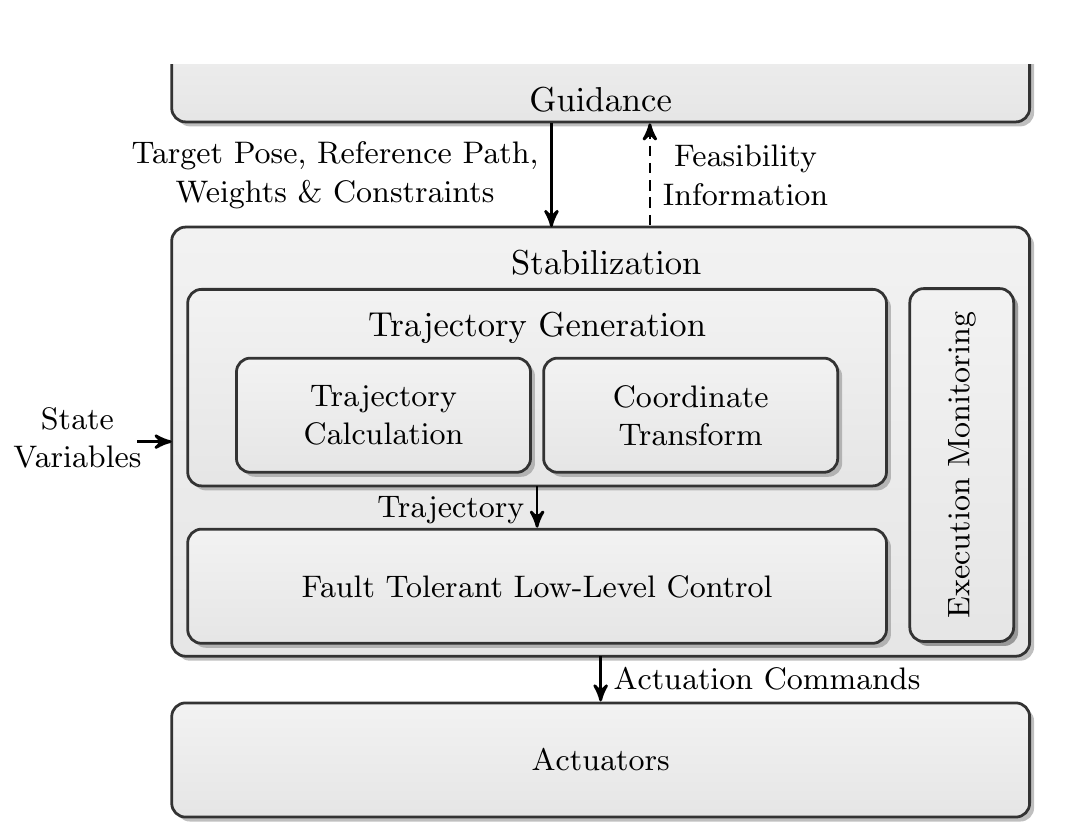}%
	\caption{Trajectory generation and control architecture, according to \cite{nolte2017}, based on \cite{matthaei2015e}. Both functional blocks apply MPC-based approaches. Execution monitoring provides degradation information at the stabilization level. \label{fig:architecture}}%
\end{figure}

The following results of the sensitivity analyses have been obtained in the framework of our research vehicle MOBILE.
MOBILE is an electric x-by-wire vehicle featuring individually steerable and drivable wheels.
On the control side, the functional architecture is separated into a trajectory generation and a trajectory control layer (cf. \autoref{fig:architecture}).
As the x-by-wire system provides no mechanical fall backs in case of actuator failures, the architecture has been designed to include fault tolerant trajectory planning and tracking modules \cite{stolte2018b, nolte2017}.
The fault tolerant control modules are designed to exploit the over-actuated actuator topology, e.g. by using torque-vectoring to compensate for steering failures.
The planning modules react to degradation by planning less dynamic trajectories to reduce demands on the actuators.

To coordinate the actuators, both, the trajectory generation and -control layers (cf. \autoref{fig:architecture}), are based on Model Predictive Control approaches \cite{stolte2018b, nolte2017}.
The trajectory generation layer generates trajectories for a horizon of several seconds, using a dynamic single-track model.
To realize fault-tolerant control, the control layer relies on a detailed double-track and a Pacejka tire model.
This separation is partially motivated by safety aspects, as the controller shall always receive a trajectory to stop on in case of an emergency.
However, an additional practical reason for the separation is given by the computational demands for the more detailed model, as addressed by \cite{polack2017}.

Because of the heavy use of model-based planning and control strategies, model quality is an important issue when judging the vehicle's performance.
For this reason, we conducted two sensitivity analyses on both models which will be described in the following.

\subsection{Double Track Model with Pacejka Tire Model}
\label{sec:nonlinear_dtm}
For the following simulations, we used a similar double track model as formulated in \cite{stolte2018b} with added roll and pitch dynamics to account for varying tire loads.
The resulting non-linear system of ODEs consists of ten state equations and 39 parameters.
In following, superscripts $W$ and $V$ denote an entity defined in the individual wheel frames and the vehicle frame, respectively (c.f. \autoref{tbl:doubletrack} for an exhaustive nomenclature for the single- and double track quantities).
\begin{table}%
	\centering%
	\caption{Nomenclature used for presented model equations. \label{tbl:doubletrack}}%
	\begin{tabular}{p{0.25\columnwidth}p{0.65\columnwidth}}%
		\toprule%
		\multicolumn{2}{l}{Super- and Subscripts}\\ \midrule%
		$(\cdot)^a, a \in \{\mathrm{V,W}\}$& vehicle or wheel coordinate frame\\
		$(\cdot)_b, b \in \{x,y\}$		   & translational or rotational quantity along/around $x$- or $y$-axis of respective frame\\
		$(\cdot)_i, i \in \{\mathrm{f,r}\}$& front or rear axle\\
		$(\cdot)_j, j \in \{\mathrm{r,l}\}$& left or right side\\ \midrule
		Inputs & \\ \midrule%
		$\delta_{i, ij}$, $\delta_{ij}$&axle-$^\ast$ or wheel individual steering angles\\
		$M^W_{ij}$& wheel individual drive torques\\
		 \midrule%
		States &\\ \midrule
		$v^\mathrm V_b$& lateral or longitudinal vehicle velocity\\
		$\omega_{ij}$& individual rotational wheel speeds\\
		$\beta$& side slip angle$^\ast$\\
		$\dot{z}_\mathrm s$& lift rate\\
		$\dot{\varphi}$&roll rate\\
		$\dot{\theta}$& pitch rate\\ 
		$\dot{\psi}$& yaw rate$^\ast$\\\midrule
		 Parameters&\\ \midrule
		$m$ & vehicle mass$^\ast$\\
		$g$ & gravitational constant$^\ast$\\
		$J^\mathrm V_\mathrm{b,z}$ & moment of inertia around the vehicle's axes$^\ast$\\
		$J^\mathrm W_{ij}$ & individual wheels' moment of inertia\\
		$l_i$ & distance between vehicle's CoG and front / rear axle$^\ast$\\
		$s_i$ & track width front / rear\\
		$h$ & height of vehicle's CoG above ground\\
		$r_{ij}$ & individual wheels' radius\\
		$d_{ij}$ & individual wheels' suspension damping factor\\
		$k_{ij}$ & individual wheels' suspension stiffness\\
		$B_a, C_a, D_a,E_a$, $S_{ij}$ & Pacejka Magic Tire Formula coefficients\\
		$c_{\alpha, i}$ & cornering stiffness per axle$^\ast$\\ \midrule
		\multicolumn{2}{l}{Other Quantities}\\\midrule
		$F^\mathrm V_{b}$& resulting forces at vehicle's CoG\\
		$F^\mathrm W_{b,ij}$& resulting forces at wheels\\ 
		$\lambda_{ij}$& longitudinal slip\\
		$\alpha_{ij}$& lateral slip\\\midrule[0.1em]%
		\multicolumn{2}{l}{${}^\ast$: quantity of single-track model}\\
	\end{tabular}%
\end{table}%

Considering the force equilibria at the tires $F^\mathrm W_{b,ij}$ and the vehicle's center of gravity $F^\mathrm V_{b}$ yields the following state equations with $p_{ij} \in \{\frac{s_\mathrm f}{2}, -\frac{s_\mathrm f}{2}, \frac{s_\mathrm r}{2}, -\frac{s_\mathrm r}{2}\}$ and $q_{ij} \in \{l_\mathrm{f}, l_\mathrm{f}, -l_\mathrm{r}, -l_\mathrm{r}\}$
\begin{IEEEeqnarray*}{rCl}%
	\psiddot	   &=& \quad \frac{1}{J^\mathrm V_z}\sum_{ij} \left[(p_{ij} \cos{\delta_{ij}} + q_{ij} \sin{\delta_{ij}})F_{x,i}^\mathrm W \right.\\
				   &&+\>\left.(q_{ij} \cos{\delta_{ij}} - p_{ij} \sin{\delta_{ij}})F_{y,i}^W\right],\\
	\dot{v}_x^V    &=& \quad  v_y^V\psidot + \frac{1}{m}\sum_{ij} \left [\cos{\delta_{ij}} F_{x,ij}^\mathrm W - \sin{\delta_{ij}} F_{y,ij}^\mathrm W\right ],\\
\end{IEEEeqnarray*}%
\begin{IEEEeqnarray*}{rCl}%
	\dot{v}_y^V    &=&  - v_x^V\psidot + \frac{1}{m}\sum_{ij} \left [\sin{\delta_{ij}} F_{x,ij}^\mathrm W + \cos{\delta_{ij}} F_{y,ij}^\mathrm W\right ],\\
	\ddot{z}_s     &=& \quad  -g + \frac{1}{m} \sum_{ij} F_{z,ij},\\
	\ddot{\varphi} &=& \quad \frac{h}{J^\mathrm V_x} \sum_{ij} F_{y,ij}^V - \frac{1}{J^\mathrm V_x} \sum_{ij} p_{ij} F^\mathrm V_{z,i},\\
	\ddot{\theta}  &=& \quad \frac{-h}{J^\mathrm V_y} \sum_{ij} F_{x,ij}^V - \frac{1}{J^\mathrm V_y}\sum_{ij} q_{ij} F^\mathrm V_{z,i},\\
	\dot{\omega}_i &=& \quad \frac{1}{J_{ij}^W} (M_{ij}^W-r_{ij} F_x^W).
\end{IEEEeqnarray*}%
The varying tire loads $F_{z,ij}$ are given in relation to the lengths $l'_{ij} \in \{l_\mathrm r,l_\mathrm r, l_\mathrm f, l_\mathrm f\}$, stiffnesses $k_{ij} \in \{k_\mathrm f, k_\mathrm f, k_\mathrm r, k_\mathrm r\}$ and damping factors $d_{ij} \in \{d_\mathrm f, d_\mathrm f, d_\mathrm r, d_\mathrm r\}$ around the stationary tire loads
\begin{IEEEeqnarray*}{rCl}%
	F_{z,ij}&=&\frac{m g l'_{ij}}{2(l_\mathrm f+l_\mathrm r)}\\
			&&+\underbrace{d_{ij}(p_{ij}\varphi+q_{ij}\theta+z_s)+k_{ij}(p_{ij}\phidot+q_{ij}\thetadot+\dot{z}_s)}_{=:\,\Delta F_{z,ij}}.
\end{IEEEeqnarray*}%
The longitudinal and lateral wheel forces $F_{a,ij}^\mathrm W$ are calculated from a Pacejka \textit{Magic Formula Tire Model} as functions of longitudinal $\lambda_{ij}$ slips and its slip angles $\alpha_{ij}$.

The symbolic equations for the sensitivity system consisting of the $10\times39$-dimensional Jacobian $\mat{J}$ and 39-dimensional vector $\vec{f_\mathrm c}$ was calculated using \MATH\footnote{the corresponding \MATH -files can be reviewed under \url{https://cutt.ly/ZrY5c5P}}.
For the sake of brevity, we resort to extracting some of the key findings from the analysis of the sensitivities in \autoref{sec:results}.
For an explicit example, the results of the sensitivity analysis for the single track model will be presented in the following section.

\subsection{Linear Single Track Model}
\label{sec:linear_stm}
A corresponding linear single track model can directly be derived from the forces calculated for the above double track model, introducing the usual assumptions:
Assuming linear tire dynamics ($\frac{1}{\alpha_{ij}} F_{y,ij}^\mathrm W= \mathrm{const.}$), constant velocity, and small steering/slip angles, as well as neglecting the influence of pitch and roll dynamics, we describe the lateral vehicle dynamics in the linear system
\begin{IEEEeqnarray*}{rCl}%
		\left ( \begin{matrix}%
			\Dot{\beta}\\
			\Ddot{\psi}\\
			\end{matrix} \right )%
		&=&%
		\left ( \begin{matrix}%
			-\frac{c_{\alpha, \mathrm f}+c_{\alpha, \mathrm r}}{m v} & \frac{c_{\alpha, \mathrm r} l_\mathrm r - c_{\alpha, \mathrm f} l_\mathrm f}{m v^2} - 1\\
			\frac{c_{\alpha, \mathrm r} l_\mathrm r-c_{\alpha, \mathrm f} l_\mathrm f}{J^\mathrm V_z} & -\frac{c_{\alpha, \mathrm f} l_\mathrm f^2+c_{\alpha, \mathrm r} l_\mathrm r^2}{J^\mathrm V_z v}%
			\end{matrix} \right )%
		\cdot%
		\left ( \begin{matrix}%
			\beta\\
			\Dot{\psi}%
			\end{matrix} \right )\\
		&&+%
			\left(%
			\begin{matrix}%
			\frac{c_{\alpha, \mathrm f}}{m v} & \frac{c_{\alpha, \mathrm r}}{m v}\\
			\frac{c_{\alpha, \mathrm f} l_\mathrm f}{J^\mathrm V_z} & -\frac{c_{\alpha, \mathrm r} l_\mathrm r}{J^\mathrm V_z}\\
			\end{matrix} \right )%
		\cdot%
		\left ( \begin{matrix}%
			\delta_\mathrm f\\
			\delta_\mathrm r%
			\end{matrix} \right ).%
\end{IEEEeqnarray*}%

The linear tire dynamics are subsumed in the cornering stiffnesses,
\begin{equation*}%
	c_{\alpha, i} = \frac{F^W_{y,i}}{\alpha_i}.%
\end{equation*}%
To ensure model consistency between the non-linear and the linearized model, this is expressed as a Taylor expansion of the non-linear tire model around $\alpha_i = 0$ and holds for small slip angles $\alpha_i$ near zero.

Applying the sensitivity analysis defined in \autoref{sec:sensitivity} yields a sensitivity system which remains simple enough to give examples for the vehicle mass and the length from the vehicle's CoG to the front. 
For the sake of clarity, we display block matrices $\mat{J}_{m,l_\mathrm f}$ and the corresponding entries of the vector $\vec f_\mathrm c$, namely $\vec f_{\mathrm c_{m, l\mathrm f}}$:
\begin{IEEEeqnarray*}{rCl}%
			\vec{\dot{Z}}_m%
			&=&%
			\left ( \begin{matrix}%
				-\frac{\beta (-c_{\alpha, \mathrm r}-c_{\alpha, \mathrm f})}{m^2 v}-\frac{\psidot (c_{\alpha, \mathrm r} l_\mathrm r-c_{\alpha, \mathrm f} l_\mathrm f)}{m^2 v^2}-\frac{c_{\alpha, \mathrm r} \delta_\mathrm r}{m^2 v}-\frac{c_{\alpha, \mathrm f} \delta_\mathrm f}{m^2 v}\\
				0
			\end{matrix} \right )\\
			&+&%
			\left ( \begin{matrix}%
				-\frac{c_{\alpha, \mathrm f}+c_{\alpha, \mathrm r}}{m v} & \frac{c_{\alpha, \mathrm r} l_\mathrm r - c_{\alpha, \mathrm f} l_\mathrm f}{m v^2} - 1\\
				\frac{1}{J^\mathrm V_z}(c_{\alpha, \mathrm r} l_\mathrm r-c_{\alpha, \mathrm f} l_\mathrm f) & -\frac{1}{J^\mathrm V_z v}(c_{\alpha, \mathrm f} l_\mathrm f^2+c_{\alpha, \mathrm r} l_\mathrm r^2)\\
			\end{matrix} \right )\vec{Z}_{m},\\
			\vec{\dot{Z}}_{l_\mathrm f}%
			&=&%
			\left ( \begin{matrix}%
				-\frac{1}{m v^2} c_{\alpha, \mathrm f} \psidot\\
				-\frac{1}{J^\mathrm V_z} (\beta c_{\alpha, \mathrm f} + c_{\alpha, \mathrm f} \delta_\mathrm f - \frac{1}{v}2 c_{\alpha, \mathrm f} l_\mathrm f \psidot)%
			\end{matrix} \right )\\
			&+&%
			\left ( \begin{matrix}%
				-\frac{1}{m v} (c_{\alpha, \mathrm f}+c_{\alpha, \mathrm r}) &\hspace{-1ex} \frac{1}{m v^2} (c_{\alpha, \mathrm r} l_\mathrm r - c_{\alpha, \mathrm f} l_\mathrm f)-1\\
				\frac{1}{J^\mathrm V_z}(c_{\alpha, \mathrm r} l_\mathrm r-c_{\alpha, \mathrm f} l_\mathrm f) & \hspace{-1ex} -\frac{1}{J^\mathrm V_z v}(c_{\alpha, \mathrm f} l_\mathrm f^2+c_{\alpha, \mathrm r} l_\mathrm r^2)\\
					\end{matrix} \right )\vec{Z}_{l_\mathrm f}.%
\end{IEEEeqnarray*}

In the following we will provide the most interesting findings from analyzing the models described in this section.
	\section{Simulative Results}
\label{sec:results}
In the following, the sensitivity systems obtained in \autoref{sec:application} are used for two main purposes:
On the one hand, we analyze nominal sensitivities in a typical operational design domain (ODD) in which both models are applied.
This is done to identify the dominant parameters in both models under non-challenging conditions.
On the other hand, we analyze to what extend the calculated sensitivities can support a monitoring framework for estimating model quality at system runtime.

For an evaluation of typical sensitivities under nominal operation, we use reference trajectories from a dataset presented in \cite{stolte2019a}. 
To generate this set, \citeauthor{stolte2019a} analyzed a g-g diagram recorded from manual drives under nominal conditions on Braunschweig's inner city ring road.
It showed that the vast majority of tuples of $(a_x, a_y)$ is contained in an interval of $a_x,a_y \in \left[-3,3\right] \si{\meter\per\second^2}$.
Hence, for those nominal cases $a_y$ does not reach the $0.5 g$ boundary as derived in \cite{polack2017}.
For this paper, the reference trajectories have been tracked in simulation with our fault-tolerant low-level controller.
Under the assumption of an Ackermann steering geometry, the resulting steering angles of the double-track model $\delta_{ij}$ have been converted to the front- and rear steering angles $\delta_{i}$ of the single track model.

With respect to evaluating the sensitivities for use in a monitoring framework, we performed the following experiments in simulation:
First, we looked at the sensitivity systems in known edge cases for model validity to get an impression what responses to expect from the sensitivity system.
Considering the single track model, we took the contributions by \citeauthor{polack2017} as a reference.
For this, we analyzed the steady-state sensitivities of the single track model on constant circular paths.
We varied the vehicle's speed to obtain lateral accelerations around the derived critical lateral acceleration of $0.5 g$.
In addition, we also analyzed the step responses of the sensitivity system with respect to front steering angle steps, driving the vehicle to circular path with smaller radius.

For the double track model, we also introduced faults into the system (e.g. locked steering angles) to drive the tires into saturation.
In this setting, we analyzed the model's sensitivities on the resulting error trajectories while the controller tried to stabilize the vehicle.
We compared the sensitivities of both models to get a feeling for the importance of parameters in both models and discuss different dynamic behavior of both sensitivity systems.

The units of the sensitivities of a state $x_i$ with respect to the parameter $c_k$ are given as 
\begin{equation*}%
	[Z_{\.i,k}] = \frac{[x_i]}{[c_k]}.%
\end{equation*}
Note, that possible factors included in SI-units which apply to the state equations, similarly affect the scale of the sensitivities.
The numeric values of the sensitivities can hence be compared without further normalization.

The following results will be presented separately for both models, focusing on the most important findings.
All box plots are configured as follows: The median value is illustrated by the line inside the boxes, the boxes contain the upper and lower quartile of measurements (i.e. 50\% of measurements are contained in the boxes, defining the inter quartile range (IQR)).
The whiskers are configured to contain lower/upper quartile $\mp 1.5 \cdot \text{IQR}$.
The mean value is depicted by a cross, outliers are displayed as solid dots.

\subsection{Single Track Model Sensitivities}
The results of the sensitivity analysis across the trajectories in our ODD-specific dataset are displayed in the upper plot of \autoref{fig:sensitivities}.
The boxplots show the distribution of selected absolute sensitivities over all trajectories.
With absolute sensitivities $|\frac{\partial \dot \psi}{\partial l_{\mathrm{f,r}}}|$ of $\sim\SI{0.1}{(\radian\,\second^{-1})\meter^{-1}}$, the lever arms around the vehicle's center of gravity dominate the model in comparison to all other sensitivities which are lower by at least an order of magnitude.
Sensitivities with respect to the mass, which are only non-zero for side-slip angle $\beta$, and with respect to the moment of inertia $J_z^\mathrm V$ show average values of \num{e-7} to \num{e-4}.
Notably, the average sensitivities with respect to the cornering stiffnesses have an order of magnitude of \num{e-5} to \num{e-4}.
This shows that for normal operating conditions, precise knowledge of the position of the vehicle's center of gravity is more important than an exact identification of  the cornering stiffnesses.

In addition to the ODD-specific sensitivities in \autoref{fig:sensitivities}, an excerpt of the steady-state sensitivities is displayed in \autoref{tab:sens}.
\begin{table}[t]%
	\centering%
	\caption{Single track model sensitivities on circular paths with varying lateral acceleration \label{tab:sens}}%
	\begin{tabular}{lcc}%
		\toprule%
		$a_y$ in $\si[per-mode=fraction]{\meter\per\second^2}$ & $\frac{\partial \beta}{\partial c_{\alpha, \mathrm f}}$ in $\si[per-mode=fraction]{\radian\per(\newton\,\radian^{-1})}$& $\frac{\partial \dot\psi}{\partial c_{\alpha, \mathrm f}}$ in $\si[per-mode=fraction]{\radian\,\second^{-1}\per(\newton\,\radian^{-1})}$\\\midrule
		$3.0$   &\num[exponent-product = \cdot]{-7.6e-8} &\num[exponent-product = \cdot]{9.7e-7}\\
		$4.0$   &\num[exponent-product = \cdot]{-5.3e-7} &\num[exponent-product = \cdot]{4.9e-6}\\
		$\mathbf{4.9}$ &\bfseries \num[exponent-product = \cdot]{-1.1e-6} &\bfseries \num[exponent-product = \cdot]{8.9e-6}\\
		$6.0$   &\num[exponent-product = \cdot]{-2.2e-6} &\num[exponent-product = \cdot]{1.4e-5}\\ \bottomrule
	\end{tabular}%
\end{table}
Comparing the obtained sensitivities for the constant curvature maneuvers around a lateral acceleration of $0.5g$ only shows that the absolute values of the sensitivities increase with increased lateral acceleration.
While the sensitivities grow beyond the values recorded for the ODD-specific dataset, the sensitivities on the circular paths alone provide no hints of sudden changes in the behavior of the model.

A more detailed system-theoretic analysis of the dynamics of the parameter variant sensitivity system will be part of future research.
The question here is whether there are further insights to be gained for the behavior of the sensitivities of the linear single track model.
In the following, we focus on the sensitivity analysis of the double-track model. 

\subsection{Double Track Model Sensitivities}
\label{subsec:sens_model_errors}

For our ODD-specific dataset, we mainly compared the resulting sensitivities to those also available in the single track model.
As $\beta$ is only part of the measurement equation of the double track model, we compute the sensitivities of $\beta$ with respect to a parameter $c$ by using the lateral and longitudinal velocities
\begin{equation*}%
	\frac{\partial \beta}{\partial c} = \frac{\partial}{\partial c} \arctan{\left(\frac{v_y^V}{v_x^V}\right)}%
	= \frac{v_x^V \cdot Z_{v_y^V,\:c} - v_y^V \cdot  Z_{v_x^V,\:c}}%
	{\left (v_x^V\right )^2 + \left (v_y^V\right )^2}.%
\end{equation*}

\begin{figure}[b]%
	\centering%
	\begin{minipage}{.95\columnwidth}%
		\myincludegraphics{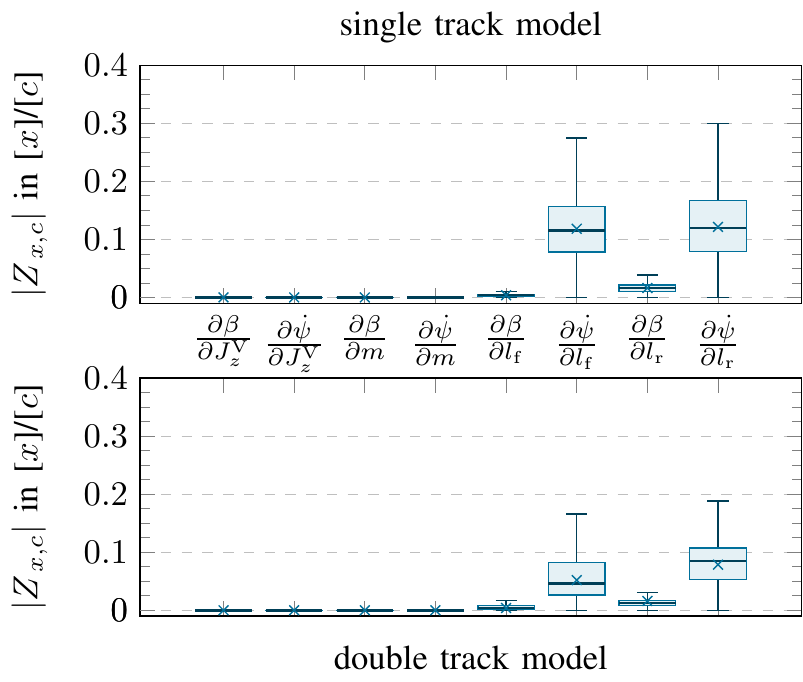}%
	\end{minipage}%
	\caption{Summary of sensitivities captured in ODD-specific dataset: single- and double track model\label{fig:sensitivities}}%
\end{figure}

A comparison of the sensitivities which can be obtained from both, the single track and the double track model, is displayed in \autoref{fig:sensitivities}.
It becomes obvious, that the sensitivities of both models show comparable behavior.
Under nominal driving conditions in the ODD, yaw rate and side slip angle are dominated by the position of the vehicle's center of gravity.
The sensitivities show comparable orders of magnitude as the ones of the single track model.

\subsection{Double Track Model Sensitivities Under Actuator Failures}

\begin{figure}%
	\centering%
	\begin{minipage}{.95\columnwidth}%
		\centering%
		\myincludegraphics{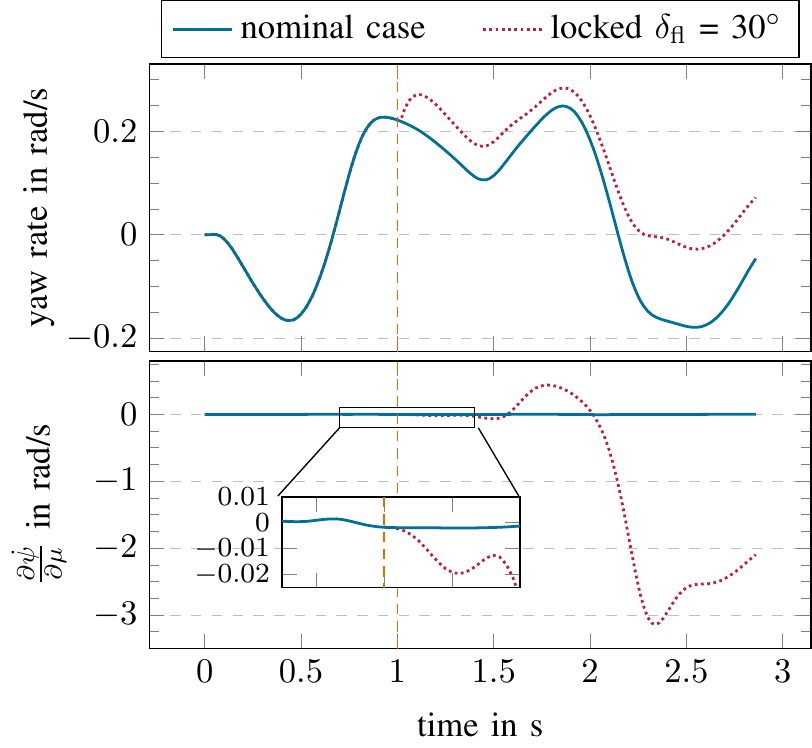}%
	\end{minipage}%
	\caption{Comparison of the resulting yaw rate, and its sensitivity with respect to $\mu$ under nominal conditions and under the influence of a locked steering actuator. Nominal case displayed in solid, blue, failure case in dotted red. Failure is introduced at $t = \SI{1}{\second}$ and significantly impacts the sensitivity. \label{fig:dt_err}}%
\end{figure}

Interesting effects show when the double track model is applied in situations at the limits of handling (cf. \autoref{fig:dt_err}).
A key driver for these experiments is the application of our double track model for fault tolerant control in case of actuator failures, as mentioned in \autoref{sec:application}.
To analyze the sensitivities under actuator failures, we performed several simulations for trajectories in the ODD-specific dataset described above.
For each trajectory, we recorded multiple simulations with an actuator fault (e.g. locked steering for one of $\delta_{ij}$, free-running, or locking wheels) induced after a given time.
The fault-tolerant controller is used to try to compensate the introduced faults, as discussed in \cite{stolte2018b}.
The states and the sensitivities for the states with respect to the model parameters have been computed for each of those simulations.
Each nominal, fault-free trajectory and the according sensitivities served as a base-case for comparison. 

\begin{figure}%
	\centering%
	\begin{minipage}{.95\columnwidth}%
		\centering%
		\myincludegraphics{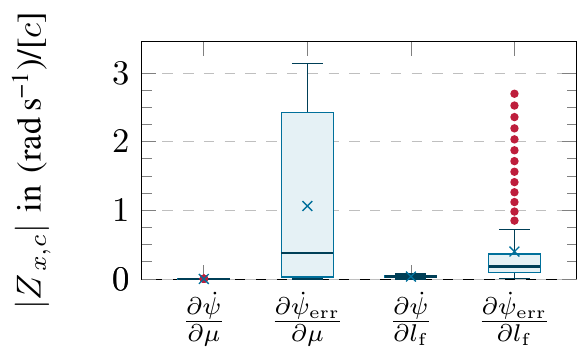}%
	\end{minipage}%
	\caption{Summary of sensitivities over the trajectories displayed in \autoref{fig:dt_err}. For comparison, influences of $\mu$ and $\l_\mathrm f$ on the yaw rate are displayed. \label{fig:dt_err_sens}}%
\end{figure}

\autoref{fig:dt_err_sens} shows a result of those simulations which includes a locked steering actuator at the front left wheel (fixed at \SI{30}{\degree}).
The upper part of the figure shows yaw rate over time for the base-case trajectory (solid, blue line), as well as the yaw rate created by the fault-tolerant low-level controller (dotted, red line).
The lower part shows the sensitivity of the yaw rate $\dot \psi$ with respect to the road-tire friction coefficient $\mu$ over time under nominal conditions (solid, blue line) and in the error case (dotted, red line).
The fault is applied at $t=\SI{1}{\second}$ (dashed, orange line).

The fault causes the tire to approach saturation.
In consequence, the front axle of the vehicle can not convey the nominally required lateral force to achieve the desired yaw rate.
The fault tolerant controller is not able to fully compensate for the actuator failure.
This results in a maximal yaw rate deviation of \SI{0.16}{\radian\per\second}.

In the following, we consider the sensitivity of the yaw rate with respect to $\mu$ (lower plot) as an example.
It is evident that there is a significant change between the nominal and the failure case.
The absolute mean sensitivity over the nominal trajectory has an order of magnitude of \num{e-4}.
When the actuator failure is induced, it rises to an order of magnitude of \num{e-1}.
The absolute maximum sensitivity rises from an order of magnitude of \num{e-3} to an order of magnitude of \num{e1}. 
This shift can be explained, as the fault causes the tire model to operate in different regions of the corresponding Pacejka tire curve:
The tire model's operating point is moved from the linear region, which dominates in the fault-free case to a non-linear region.
In this case, the lateral dynamics of the double-track model are dominated by the tire parameters and accurate estimation of those parameters becomes crucial for the model to reflect the dynamics accurately.

The sensitivities of the yaw rate with respect to the lever arms (cf. \autoref{fig:dt_err_sens}) support this statement.
While these sensitivities increase in the failure case as well, the change is not as significant as for the sensitivity with respect to $\mu$.
For $Z_{\dot\psi, l_\mathrm{f}}$, the absolute mean as well as the absolute maximum sensitivity have an order of magnitude of \num{e-2}.
In the failure case, the absolute mean rises to an order of magnitude of \num{e-1}, the absolute maximum rises to an order of magnitude of \num{e1}.

The resulting distributions of the sensitivities support these findings, as well.
While there are comparable absolute maximum values of the sensitivities of $\dot{\psi}$ with respect to $\mu$ and $l_\mathrm f$, the boxplots show them as outliers in the case of $l_\mathrm f$.
The inter quartile ranges in the failure case show a significantly wider distribution of the sensitivity with respect to $\mu$ and a significant amount of the upper quartile range having magnitudes of \num{e1}.
The box for the sensitivity with respect to $l_\mathrm f$ stays significantly smaller than 1.

These quantitative findings support the intuitive impression that the tire characteristics dominate the model's behavior at the limits of handling.
At the same time, from a monitoring perspective, these findings indicate that sensitivities can support online diagnosis by giving additional information about the influence of model parameters at runtime.
	\section{Conclusion and Future Work}%
\label{sec:conclusion}%
In this paper we have argued the value of a model theoretic discussion of applied vehicle dynamics models for automated driving.
We pointed out the importance of being critical of and gaining as much information about the applied models in model-based algorithms, particularly with respect to safety assurance for automated driving systems.
Knowledge about the utilized models is more critical when (optimal) feed-forward control approaches are applied - even more so under longer prediction horizons, when no direct feedback from measured states can be applied to compensate for model inaccuracies.

We gave an example of sensitivity analysis for a single and double track vehicle dynamics model as an application of typical tools from modeling theory.
This analysis was conducted for trajectories from an urban ODD, as well as near known model boundaries.
The results we obtained for the sensitivities of both models showed that in nominal driving conditions, the models' behavior is dominated by the location of the position of the vehicle's center of gravity.
The analysis of the double track model's sensitivities gave quantitative proof of the fact that model behavior in dynamically challenging situations is dominated by the tire parameters.
This changing sensitivity of the model is a major concern for the application of model-based approaches for fault-tolerant control.
Without proper awareness of which parameters dominate a model at given operating points, the application of model-based algorithms in this domain might be infeasible.

For future work, we aim at formal analyses of the sensitivity systems in order to allow more general statements about the dynamics of the sensitivity systems.
A further line of research will be the integration of the sensitivity systems into a monitoring framework to allow conducting runtime analyses of vehicle dynamics models to contribute to increased autonomy of automated vehicle systems.
	\IEEEtriggeratref{7}
	\renewcommand*{\bibfont}{\footnotesize} 
	\printbibliography 
\end{document}